# Prediction of Human Empathy based on EEG Cortical Asymmetry

1st Andrea Kuijt
*dept. of Cognitive Science and AI Tilburg University*
Tilburg, Netherlands
andreakuijt@gmail.com

2nd Maryam Alimardani
*dept. of Cognitive Science and AI Tilburg University*
Tilburg, Netherlands
m.alimardani@uvt.nl

*Abstract*—Humans constantly interact with digital devices that disregard their feelings. However, the synergy between human and technology can be strengthened if the technology is able to distinguish and react to human emotions. Models that rely on unconscious indications of human emotions, such as (neuro)physiological signals, hold promise in personalization of feedback and adaptation of the interaction. The current study elaborated on adopting a predictive approach in studying human emotional processing based on brain activity. More specifically, we investigated the proposition of predicting self-reported human empathy based on EEG cortical asymmetry in different areas of the brain. Different types of predictive models i.e. multiple linear regression analyses as well as binary and multiclass classifications were evaluated. Results showed that lateralization of brain oscillations at specific frequency bands is an important predictor of self-reported empathy scores. Additionally, prominent classification performance was found during resting-state which suggests that emotional stimulation is not required for accurate prediction of empathy -as a personality trait- based on EEG data. Our findings not only contribute to the general understanding of the mechanisms of empathy, but also facilitate a better grasp on the advantages of applying a predictive approach compared to hypothesis-driven studies in neuropsychological research. More importantly, our results could be employed in the development of brain-computer interfaces that assist people with difficulties in expressing or recognizing emotions.

## I. INTRODUCTION

One of the challenges in the field of human-technology interaction that has long received attention is to endow a machine with the ability to understand and react to a user's emotions. Many works in affective computing rely on analysis of behaviour modalities such as facial expressions, bodily gestures and speech to perceive and interpret the user's emotional states [1]. However, with the advancement of sensing technology, recent studies are leaning toward emotion recognition based on (neuro)physiological data [2] as they can pick up unconscious affect changes and also be used as a proxy for those who suffer from communication and behavior disorders such as autism.

Emotion recognition based on neuroimaging approaches, besides providing fundamental knowledge about cognitive functions of the brain, has some practical implications. The outcomes can be used for development of brain-computer interfaces that enhance the interaction between human and technology. In particular, machines that can detect and react to our stress, confusion or enjoyment levels would be able to improve our productivity at work, facilitate learning or maximize entertainment [3]. Artificial agents that are able to predict human affective state would give rise to virtual diagnosing of social behaviour disorders [4] while assisting these individuals on how to deal with their condition. For instance, virtual coaches can adapt their behaviour to the personality of the user, e.g. verbal encouragement [5], or assist people with difficulties to express or recognize emotions [6]. In addition, affective adaptation holds the potential of boosting the efficacy of virtual tutors [7]. As emotions are strongly linked to attention, motivation and learning, a tutor that is able to minimize boredom and frustration and to react to confusion would help users to learn faster than another tutor that only relies on performance measures [3].

An important aspect of emotional regulation is empathy. Empathy is the ability to vicariously experience and understand the affect of other people and is fundamental to successful social-cognitive functioning and behaviour [8]. Empathy entails both cognitive empathy, a good understanding of what another person is feeling in the absence of any noticeable affects [9], as well as affective empathy, the emotional reaction to the emotional state of another person [10].

Even though previous research has shown several insights, the neuroscientific investigation of empathic responses and traits is still in its early stages [11]. Frequency band analyses on electroencephalography (EEG) signals have contributed to the general understanding of the neurological basis of empathy [12]. Due to the millisecond temporal resolution of EEG, power peaks reflecting brain oscillations in specific frequency bands can be explored [13], [14]. Past research has revealed a prominent role of pre-frontal cortical asymmetry, which is the relative difference in activation between the hemispheres, in mediating empathy-related behaviors: a stronger right frontal activation seems to be associated with negative emotions, while a relatively stronger left frontal activation is associated with positive emotions [15], [16]. This is in line with the valence hypothesis [17], which states that the left hemisphere is specialized for processing of positive emotions and the right hemisphere is specialized for processing of negative emotions. More specifically, short-term changes of pre-frontal alpha asymmetry are considered to correlate with the relative activation of two motivational systems [18]: left-over-right alpha asymmetry has been linked to approach-based emotion and motivation [19] as well as affective perspective taking and cognitive empathy [20], while right-over-left alpha asymmetry is linked to withdrawal or avoidance-based responses [19].

Other studies have shown that negative situations increase brain activity within the low-frequency bands, i.e. theta and delta, of the right pre-frontal cortex, whereas positive situations elicit greater responses within the low-frequency bands of the left pre-frontal cortex [21]. Moreover, prolonged visual emotional stimulation seems to increase theta band power responses [21]. Also an association between high-frequency bands, i.e. beta and gamma, in either the frontal, central or parietal area of the brain, with positive and negative emotions was found [22]. The activation of the central-parietal regions was more salient over the left than right hemispheres, as a response to pictures of hands in painful situations [23]. Although the modulation of different frequency bands seems to significantly contribute to the explanation of arousal effects on emotional cue comprehension, the specific role of brain oscillations in affective and empathic behavior is partially unknown. As cortical asymmetry has proved to be a stable and reliable measure of individual differences [24], and proneness to feeling sadness translates into a stronger response to sadness of others [25], cortical asymmetry seems as a promising measure for prediction of empathy.

Despite the fact that previous research showed a distinct association between cortical asymmetry and human empathy [26], it is not clear whether it is possible to predict human empathy based on the same EEG components. Most of the existing research on EEG correlates of personality traits has been conducted in a hypothesis-driven way [27], without giving any serious consideration to a predictive approach. However, a common problem in neurophysiological research is the inability to gain similar results when repeating a study with a new group of participants [28].

Therefore, an alternative research approach might be more fruitful: research in which 'success' is measured not by the size of a theoretically privileged regression coefficient or a model fit statistic, but instead simply by the average difference between unobserved data (i.e., 'out-of-sample' data that were not used to fit the model) and the model's predictions for those data [29]. However, until now, research focused on the predictions of personality traits and emotional responding related to empathy based on the analysis of EEG data has not reached consistent conclusions: some papers reported a successful analysis [30], while others did not [29].

The goal of the current study is twofold; (1) it attempts to push forward the usage of EEG-analysis in predicting and understanding human emotional processing, and (2) it aims to contribute to the predictive approach in neurophysiological research by applying machine learning algorithms to accommodate for the complexity of the recorded data and to guarantee a higher degree of generalizability. EEG data collected from 52 participants who watched an immersive video of a child exploitation scene in a virtual reality headset were used. Since empathy is considered both as a personality trait and a state response, we employed our approach before the subjects were engaged in emotional processing (baseline), during the emotional stimulation (video) and after that. Therefore, the current study elaborated on the following three research questions:

1) whether EEG cortical asymmetry extracted from different brain regions in different frequency bands can predict self-reported empathy levels?

2) whether EEG cortical asymmetry extracted from different brain regions in different frequency bands can classify self-reported empathy levels into two or three subgroups?

3) whether there are any substantial differences in the association between EEG signals and self-reported empathy scores, before, during or after exposure to emotional stimulation?

Based on the research discussed above, we expected a predictive value of cortical asymmetry in human empathy, specifically in the alpha band. Moreover, we expected an influence of exposure to emotional stimulation on this relationship.

## II. METHOD

### A. Data

The data was collected by a team of researchers led by the second author [26]. Fifty-two participants watched an emotional 360° virtual reality (VR) video of a young African girl, who was being abused as a domestic slave. EEG signals were collected before, during and after the stimulus, from the frontal, central and occipital areas of the brain (9 electrodes: F3, Fz, F4, C3, Cz, C4, P3, POz and P4). This resulted in three data segments: Pre-video segment, Video segment and Post-video segment. The EEG signals were bandpass filtered between 0.5 and 50 Hz. For every subject, an array of the mean EEG power was computed for five frequency bands: delta (0.5-4 Hz), theta (4-8 Hz), alpha (8-13 Hz), beta (13-30 Hz), and gamma (above 30 Hz), in every electrode and every segment. The asymmetry powers were obtained by subtracting the logarithmic power of the left hemisphere from the logarithmic power of the right hemisphere (i.e. F4 – F3, C4 – C3, and P4 – P3). A positive asymmetry value equaled stronger activation in the right hemisphere compared to the left hemisphere. Data of the participants with outlying logarithmic cortical asymmetry values ($Z > 3.0$) were excluded.

The Toronto empathy questionnaire [31] was used to quantify the level of empathy. The questionnaire applied 7-point Likert-scale that corresponded to various levels of frequency (i.e., never, almost never, rarely, sometimes, often, very often, always), assuming that distances between each answer option are equal. The possible range of self-reported empathy scores was 0 to 96. For the participants of the current study, the obtained self-reported empathy scores ranged between 49 to 86. The self-reported empathy scores of subjects were collected before the experiment [26].

*B. Feature selection*

In total, there were 15 features extracted from the EEG data (i.e. asymmetry values of five frequency bands in three brain regions). Several feature selection methods were implemented to create a feature subset. For both the multiple linear regression models as well as for the classification models, the following feature selection methods were applied:
1. First, Linear Model Feature Ranking was implemented. This method uses three different Scikit-Learn linear models (Linear, Lasso and Ridge Regression) to create feature importance rankings.
2. Second, Recursive Feature Elimination (RFE) was used. This method uses a linear regression to select the worst-performing feature and then excludes this feature. The loss of the classification margin is used as an objective function to evaluate the discriminative contributions of each feature. Features that possess the lowest contribution are iteratively eliminated from the training set. This method elicits rankings, ordering the features from salient to non-salient [32].
3. Third, Random Forest Feature ranking was applied, using the Random Forest's attribute 'feature importances' to calculate and ultimately rank the feature importance. To combine the methods, they were all integrated into one matrix, and were used to create factor plots [33].

Moreover, in order to get a better grip on the relationship between the features and the dependent variable (i.e. empathy scores), correlations were measured between the best ranked features and the dependent variable. The p-values showed whether features were significantly correlated with the dependent variable.

*C. Multiple Linear Regression*

As the self-reported measures were continuous data, a multiple linear regression analysis was used to look into the possibilities of predicting continuous data. Two different models were compared for each segment: 1) a model with all 15 features; 2) a model with a subset of five features selected by the earlier described feature selection methods. Before the multiple linear regression models were applied, the key assumptions of a multiple linear regression were checked and confirmed, concerning the assumption of normally distributed residuals, the assumption of homoscedasticity, and the assumption of absence of multicollinearity [34]. In order to evaluate the multiple linear regression models, Mean Squared Error (MSE), Mean Absolute Error (MAE) and significance were used as evaluation metrics. MSE measures the average of the squares of the errors -that is, the average of squared difference between the estimated values and the observed values. MAE is the average of the absolute difference between the predicted values and observed values. The MAE measures the average magnitude of the errors in a set of predictions, without considering their direction. Because the difference is squared in the MSE, it is almost always larger than the MAE.

To systematically evaluate the models on multiple subsets of the dataset, the resampling method k-fold cross-validation (k=5) was used.

*D. Binary and Multiclass Classification*

The self-reported empathy scores were separated into two balanced subgroups for the binary classification, and into three unbalanced subgroups for the multiclass classification. Discretization of continuous features reduces the impact of small fluctuations in the data on the model by splitting the data into bins, and has frequently been used in studies classifying personality traits [27], [32], [35], [36], [37]. In the binary classification, a median split divided the empathy scores into two classes: 'high' vs 'low' empathy groups. In the multiclass classification, equal intervals between the minimum-maximum range was implemented. This tested the possible influence of the narrow range between 49 and 86 (i.e. the min and max of obtained empathy scores). For these classifications, three models i.e. a Logistic Regression (LR), a Support Vector Machine (SVM), and a Decision Tree (DT) were implemented using a selection of five best performing features. Table I shows the hyperparameters that were tuned in preparation. In order to evaluate the binary and multiclass classification models, the F1-score was used as an evaluation metric. The F1-score is the harmonic mean of precision and recall. A high F1-score means that the model produces few false positives and few false negatives. Moreover, F1-score is a useful metric when there are imbalanced classes as it was the case in this study. To systematically evaluate the models on multiple subsets of the dataset, the resampling method leave-one-out cross-validation was used.

TABLE I
HYPERPARAMETERS AS A RESULT OF A GRID SEARCH FOR LOGISTIC REGRESSION (LR), SUPPORT VECTOR MACHINE (SVM) AND DECISION TREE (DT), TUNED FOR THE CORTICAL ASYMMETRY FEATURES, FOR EVERY SEGMENT SEPARATELY.

| Model | Parameter tuning | | |
|---|---|---|---|
| | *Parameter* | *Tuning values* | *Best value* |
| LR | C | [1.0, 1.5, 2.0 ,2.5] | 2.0 |
| | dual | [True, False] | True |
| | max_iter | [100, 110, 120, 130, 140] | 100 |
| SVM | C | [0.001, 0.01,0.1] | 0.1 |
| | gamma | [0.001, 0.01, 0.1, 1] | 0.1 |
| | kernel | [linear, poly] | poly |
| | degree | [1,2,3,4,5] | 5 |
| DT | max_features | [auto, sqr, log] | auto |
| | min_samples_leaf | [1,2, ... 14,15] | 10 |
| | min_samples_split | [2,3, ... 14,15] | 6 |
| | criterion | [entropy, gini] | gini |

III. RESULTS

*A. Multiple Linear Regression*

When using the model with all fifteen features to predict empathy, the results showed that this model only significantly predicted the empathy scores in the Post-video segment (see Table II). When using the model with a selection of five features, the model only significantly predicted the empathy scores in the Pre-Video segment. Moreover, when considering the contribution of selected features to the prediction (see Table III), the results indicated a significant main effect for frontal alpha asymmetry in the Pre-video segment and frontal gamma asymmetry in the Post-video segment.

TABLE II
MULTIPLE LINEAR REGRESSION MODELS PREDICTING CONTINUOUS SELF-REPORTED EMPATHY SCORES, FOR PRE-VIDEO, VIDEO, AND POST-VIDEO SEGMENTS, WITH 15 FEATURES, AND WITH 5 FEATURES (AFTER FEATURE SELECTION)

|  | Evaluation results | | | |
|---|---|---|---|---|
| Model | Segment | MSE | MAE | p |
| 15 features | Pre-video | 51.749 | 6.735 | 0.583 |
|  | Video | 122.332 | 9.206 | 0.976 |
|  | Post-video | 150.556 | 10.231 | **0.045** |
| 5 features | Pre-video | 78.850 | 8.740 | **0.046** |
|  | Video | 77.876 | 7.344 | 0.407 |
|  | Post-video | 62.102 | 6.875 | 0.057 |

TABLE III
MULTIPLE LINEAR REGRESSION MODELS FOR THE PRE-VIDEO, VIDEO AND POST-VIDEO SEGMENTS AFTER FEATURE SELECTION

|  | OLS Regression Results | | | |
|---|---|---|---|---|
| Segment | Selected features | coef | t | p |
| Pre-video | central delta | 0.315 | 0.191 | 0.851 |
|  | parietal delta | 3.633 | 1.640 | 0.115 |
|  | frontal theta | 1.479 | 0.889 | 0.384 |
|  | frontal alpha | -10.748 | -3.021 | **0.006** |
|  | central gamma | 0.455 | 0.766 | 0.452 |
| Video | frontal delta | 4.430 | 1.662 | 0.111 |
|  | parietal delta | -6.252 | -1.524 | 0.142 |
|  | frontal theta | -4.540 | -1.941 | 0.066 |
|  | parietal theta | 5.942 | 1.316 | 0.202 |
|  | parietal gamma | 2.899 | 0.630 | 0.536 |
| Post-video | frontal delta | 1.521 | 0.965 | 0.345 |
|  | frontal theta | -2.354 | -0.868 | 0.395 |
|  | central alpha | 1.233 | 0.463 | 0.648 |
|  | frontal alpha | -6.909 | -1.750 | 0.095 |
|  | frontal gamma | 2.414 | 2.833 | **0.010** |

*B. Binary and multiclass classification*

For the binary classification, a model with a selection of five features was used to predict empathy (see Table IV). The highest scores were found during the Pre-video segment, in which the SVM classifier performed the best (F1-score = 0.722), and during the Post-video segment, in which the DT classifier showed the highest performance score (F1-score = 0.701).

Additionally, a multiclass classification model was implemented. The continuous empathy scores were split up into three classes. Again, a model with a selection of five features was used to predict empathy (see Table V). The DT classifier in the Post-video segment showed the highest classification performance (F1-score = 0.621).

When comparing the performance of the binary classification models with the performance of the multiclass classification models, better results were found for the binary classification than for the multiclass classification. These results suggested that dividing the continuous self-reported empathy scores into more than two classes is not preferable for our dataset.

TABLE IV
BINARY CLASSIFICATION METRICS, FOR PRE-VIDEO, VIDEO, AND POST-VIDEO SEGMENTS RESPECTIVELY, USING A LOGISTIC REGRESSION (LR), SUPPORT VECTOR MACHINE (SVM) AND DECISION TREE (DT) WITH THE FIVE BEST PERFORMING FEATURES.

|  | Classification report (classes = 2) | | | | |
|---|---|---|---|---|---|
| Segment | Model | Accuracy | Precision | Recall | F1-score |
| Pre-video | LR | 0.578 | 0.599 | 0.567 | 0.574 |
|  | SVM | 0.683 | 0.674 | 0.785 | **0.722** |
|  | DT | 0.513 | 0.444 | 0.393 | 0.412 |
| Video | LR | 0.443 | 0.443 | 0.474 | 0.461 |
|  | SVM | 0.446 | 0.453 | 0.533 | 0.492 |
|  | DT | 0.504 | 0.504 | 0.595 | 0.541 |
| Post-video | LR | 0.561 | 0.553 | 0.653 | 0.591 |
|  | SVM | 0.593 | 0.574 | 0.764 | 0.652 |
|  | DT | 0.653 | 0.614 | 0.822 | **0.701** |

TABLE V
MULTICLASS CLASSIFICATION METRICS (CLASSES = 3); FOR PRE-VIDEO, VIDEO AND POST-VIDEO SEGMENTS RESPECTIVELY, USING A LOGISTIC REGRESSION (LR), SUPPORT VECTOR MACHINE (SVM) AND DECISION TREE (DT) WITH THE FIVE BEST PERFORMING FEATURES.

|  | Classification report (classes = 3) | | | | |
|---|---|---|---|---|---|
| Segment | Model | Accuracy | Precision | Recall | F1-score |
| Pre-video | LR | 0.432 | 0.422 | 0.432 | 0.429 |
|  | SVM | 0.431 | 0.401 | 0.433 | 0.372 |
|  | DT | 0.341 | 0.301 | 0.345 | 0.322 |
| Video | LR | 0.442 | 0.391 | 0.443 | 0.391 |
|  | SVM | 0.471 | 0.371 | 0.472 | 0.389 |
|  | DT | 0.441 | 0.342 | 0.443 | 0.372 |
| Post-video | LR | 0.501 | 0.501 | 0.501 | 0.504 |
|  | SVM | 0.412 | 0.464 | 0.412 | 0.311 |
|  | DT | 0.622 | 0.633 | 0.624 | **0.621** |

IV. DISCUSSION

The aim of this study was to examine the prospect of applying a predictive approach in studying the association between cortical asymmetry and self-reported empathy. Based on previous literature [15], [17], [18], [25], [26], the expectation was that it would be possible to predict human empathy based on EEG asymmetry values, obtained in five frequency bands from different regions of the brain. Several methods were implemented in order to study this hypothesis including prediction of a continuous empathy score as well as binary and multiclass classificaitons of empathy groups. Moreover, the study was interested in whether the prediction power of the models was different when they were trained using EEG features of the resting-state, during and after participants were exposed to an empathy inducing video stimulus.

Based on the multiple linear regression analysis, a prominent role of frontal alpha asymmetry was found for the models in the Pre-video and Post-video segments. This was in line with the expectation based on previous literature [12], [15], [16], [18], [19], [25]. In addition, when the model was trained with Post-video EEG data, significant main effects were found for cortical asymmetry in high-frequency bands, i.e. gamma of the frontal cortex. This is in line with the previously reported role of high-frequency bands in providing discriminative information for emotion recognition [22]. None of the models in the Video segment showed high performance in prediction of self-reported empathy scores. Therefore, these results propose that empathy as a personality trait can be predicted based on EEG cortical asymmetry in the absence of emotional stimulation. They also suggest that future research should consider asymmetry changes from resting-state to the stimulus phase per individual rather than using EEG responses in the stimulus phase as features for classification.

Even though acceptable performance was obtained in the binary classification of empathy groups, the large prediction errors in the regression analysis showed that the difference between the estimated values and the observed values was large. The inability of these models to predict continuous empathy values can perhaps be explained by the fact that the target variable had a narrow range and the dataset used for training of the model was small. A narrow target variable range means that the model has to be particularly precise in order to gain high prediction performance and a small dataset results in a model that is highly susceptible to errors. Future work should make an attempt to replicate this study with more EEG data that is collected from a broader range of participants that have been screened for their self-reported empathy scores prior to the experiment.

Our results in the binary classification showed that a SVM model with a subset of selected features in the Pre-video segment (baseline measurement) and a decision tree model with a subset of selected features in the Post-video segment are capable of achieving a relatively desirable classification performance (F1-score > 0.7). This is in contrary to the results of a previous study, in which machine learning models were employed to predict personality traits based on power spectra of resting-state EEG but failed in achieving this task [27]. An explanation could be that the mentioned study only focused on spectral power measures as training features and did not consider predictive power of other EEG components such as cortical asymmetry. Another reason could be that this study looked at Big Five personality traits as dependant variables. Although past psychological studies have shown weak relationships between empathy and Big Five personality dimensions [38], the two are not necessarily the same.

An interesting finding in our study was the fact that frontal alpha asymmetry was selected as an important feature in both Pre-video and Post-video segments. This suggested that frontal alpha asymmetry plays an important role in predicting human empathy, which is in line with previous findings [15], [16]. Tullett et al. found that frontal alpha asymmetry was a significant predictor for empathic concern, in response to charity images [25]. On the contrary, the current study showed that it is possible to predict empathy based on frontal alpha asymmetry in resting-state. This could mean two things; either the immersive VR video that was shown to the participants of this study was not stimulating enough to induce any strong asymmetry values, or empathy as a self-reported trait can only be predicted by resting-state brain activity, without emotional stimulation.

Even though previous studies have demonstrated the possibility of recognizing human personality using a small multimodal dataset [37], the small size of the EEG dataset was an important limitation in this predictive research. Employing machine learning algorithms with small datasets results into lower predictive and classification performance, as well as a higher risk of overfitting and therefore lower generalizability. In line with the limitation regarding the small dataset, the selection and ranking of the variables are data- and algorithm-dependent. Thus, the variables selected for the employed algorithms should be viewed as one of the many possible selections and not the only one. It is therefore suggested to look at other features that are extractable from the EEG signals to be incorporated in the training of the models. Moreover, in future studies, the potential influence of gender on the relationship between cortical asymmetry and empathy should be considered. Previous literature found that women in general show greater right-dominant frontal alpha asymmetry than men [39].

In sum, the possibility of observing and predicting human empathy on the basis of neural activity introduces a new method for emotion recognition and personality profiling. This approach can be used in providing support to individuals who struggle with social behavioural disorder that is fully adapted to their condition. Also, it could be used for enhancement of human-technology interaction, where artificial agents or systems are able to understand human affective state and personalize the interaction [4].

## V. CONCLUSION

This study elaborated on adopting a predictive approach in studying the neural bases of human empathy. Multiple machine learning algorithms were employed to predict empathy levels of humans based on EEG asymmetry values in five frequency bands (delta, theta, alpha, beta, and gamma) and three brain regions (frontal, central and parietal). Although multiple linear regression model failed to predict accurate values of continuous empathy scores, promising results were found in the binary classification of 'high' vs. 'low' empathy groups. In almost all models, EEG signals that were recorded during resting-state i.e. before and after participants were emotionally stimulated, seemed to be better predictors than EEG signals during the emotional stimulation. Also a significant contribution was found for frontal alpha asymmetry in predicting empathy as a personality trait. Our results hold promise for development of future brain-computer interfaces that predict a user's personality traits and emotional responses to improve the quality of interaction during technology usage.

## ACKNOWLEDGMENT

The current research was partially funded by the European Union, OP Zuid, the Ministry of Economic Affairs, the Province of Noord-Brabant and the municipalities of Tilburg and Gilze Rijen (PROJ-00076). Authors would like to thank Annabella Hermans for assisting in the data collection.